\documentclass[doublecol]{epl2}
\usepackage{amsmath}
\usepackage{amsfonts}
\usepackage{graphicx}
\newcommand{\nextnearsites}[1]{\langle\!\langle #1 \rangle\!\rangle}
\newcommand{\nearsites}[1]{\langle #1 \rangle}
\newcommand{\ii}{\mathrm{i}\,}

\title {Exactly solvable 2D topological Kondo lattice model}
\author{Igor N.Karnaukhov \and Igor O.Slieptsov}
\institute{G.V. Kurdyumov Institute for Metal Physics, N.A.S. of Ukraine, 36 Vernadsky Boulevard, 03142 Kiev, Ukraine}
\abstract {
A spin-$\frac{1}{2}$ Kitaev sublattice interacting with a subsystem of spinless fermions is studied on a honeycomb lattice when the fermion band is half filled. The model Hamiltonian describes a topological Kondo lattice with the Kitaev interaction, it is solved exactly by reduction to free Majorana fermions in a static $\mathbb{Z}_2$ gauge field. An yet unsolved problem of a hybridization of fermions and local moments in the Kondo lattice at low temperatures is solved in the framework of the model proposed. The Kondo hybridization gap is opened and the system is fixed in  insulator and spin insulator states, due to a spin-fermion nature of the gap.
We will show that the hybridization between local moments and itinerant fermions should be understand as hybridization between corresponding Majorana fermions of the spin and charge sectors.
The RKKI interaction between local moments is not realized in the model, a system demonstrates a `quasi Kondo' scenario of behavior with realization chiral gapless edge states in topological nontrivial phases.
The ground-state phase diagram of the interacting subsystems calculated in the parameter space is rich.
}

\begin{document}
\maketitle


\section{Introduction}

The Kondo lattice problem remains an unsolved problem of strong correlated systems, with interacting delocalized and localized electrons~\cite{ts,k,k0}.
The Kondo lattice Hamiltonian describes conduction electrons interacting with local moments arranged regularly. The Kondo lattice approximation is used for description of rare-earth and transition compounds.
The scenarios of hybridization of electrons and local moments, opening of spin and charge gaps at low temperatures, a formation of large volume of the Fermi surface in the Kondo lattice are still unclear. In the framework of 1D model of the Kondo lattice Tsvelik has shown that the insulating state forms not due to a hybridization of conduction electrons with local moments, but as a result of strong antiferromagnetic fluctuation~\cite{ts}.
Note, that the insulator phase in the state of the Kondo insulator is realized in all compounds with odd number of itinerant electrons per a local moment.

The Kondo lattice problem is not reduced to a single impurity Kondo problem where a local moment is screened at low temperatures by conduction electrons and the ground state of an impurity is a spin-singlet~\cite{r,r1}. Behavior of the system (the Kondo problem) is defined by the exponentially small scale (the Kondo temperature) in the exchange coupling constant.

We hope to explain the phenomenon of the Kondo lattice in the framework of the model of a topological Kondo lattice proposed. We will call a complex system, which includes interacting spin and fermion subsystems one of them or two are topological, a topological Kondo lattice. This class of compounds is now known as topological Kondo insulators. Recent experiment on $\chem{SmB_6}$ shown that a phase state of a topological Kondo insulator is realized in $\chem{SmB_6}$.
The compound $\chem{SmB_6}$  attracted now attention due to anomalous behavior of a residual conductivity at low temperature $T \sim 5-7 K$ that is characterized by a 3D to 2D crossover of the transport carriers.
The intermetallic compound $\chem{SmB_6}$ is a strongly correlated insulator with topological states at low temperatures, it is the first candidate of a topological Kondo insulator family~\cite{rev,ex1,ex2,ex3,ex4,ex5}. According to refs.~\cite{th,th1,th2,th3}, the topological state in $\chem{SmB_6}$ is a result of indirect interaction of $5d$-states via strong spin-orbit coupling and hybridization between itinerant $5d$- and narrow (localized) $4f$- electronic states.
The topological Kondo insulator is formed via strong interaction of dispersive $d$- and non-dispersive $f$-electronic bands and hybridization between them.
Describing a non-dispersive band in the framework of the spin operators the system of strongly correlated fermions can be defined on the topological Kondo lattice and the phase of a topological Kondo insulator is realized  as its the topological state.

We study a 2D model of the topological Kondo lattice at the half-filling and small doping.
The model of topological Kondo insulator is the first example of exact solvable model of a topological Kondo lattice,
the exact solution of the Kondo-like lattice model can also be crucial for understanding of the behavior of the Kondo lattice. In the framework of the model we show that the hybridization gaps arise in the state of Kondo insulator.
The physics of the  model is surprisingly rich.

\section{Model}
\begin{figure}[tbp]
\centering{\leavevmode}
\includegraphics[width=7cm]{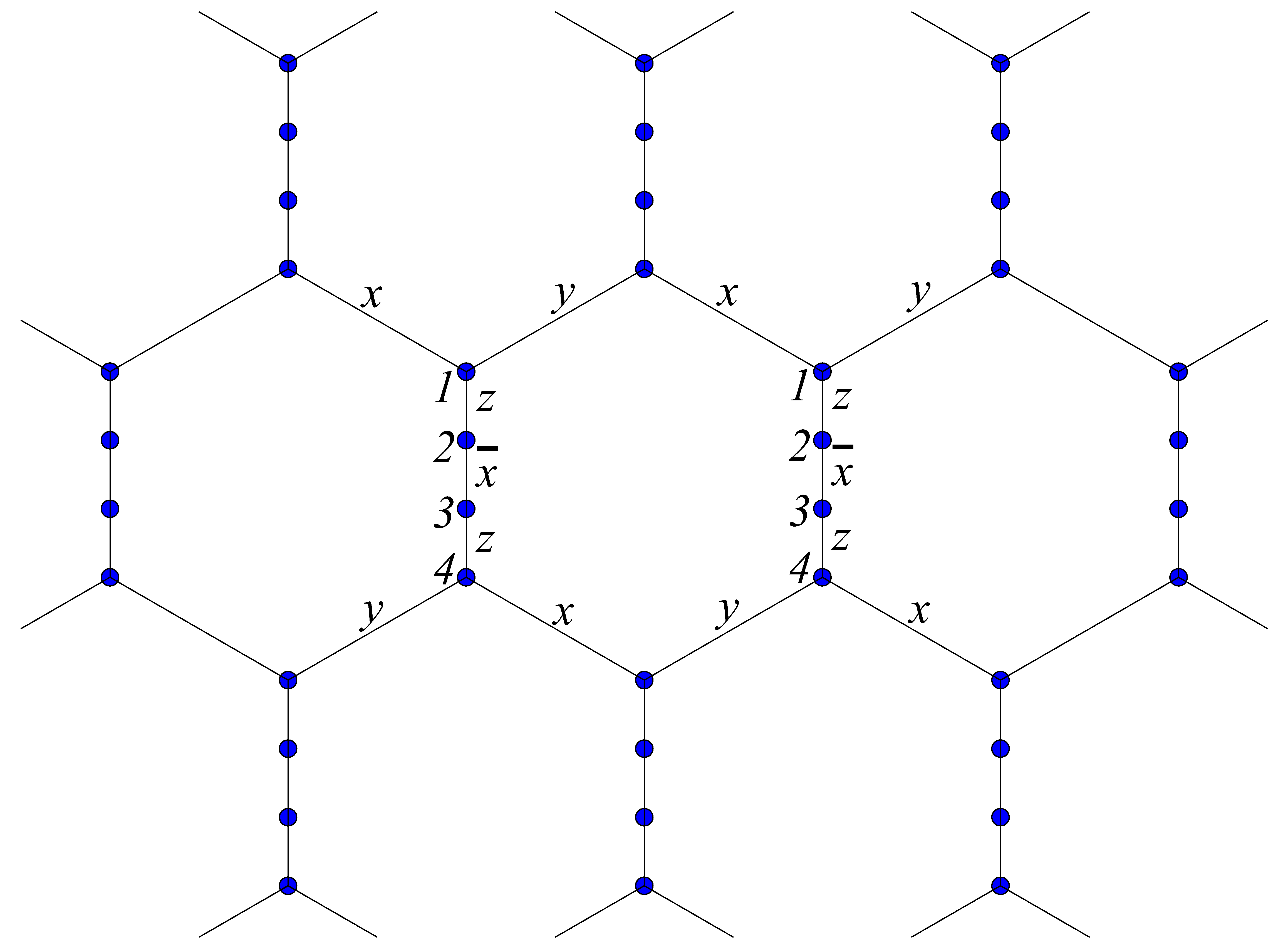}
\includegraphics[width=7cm]{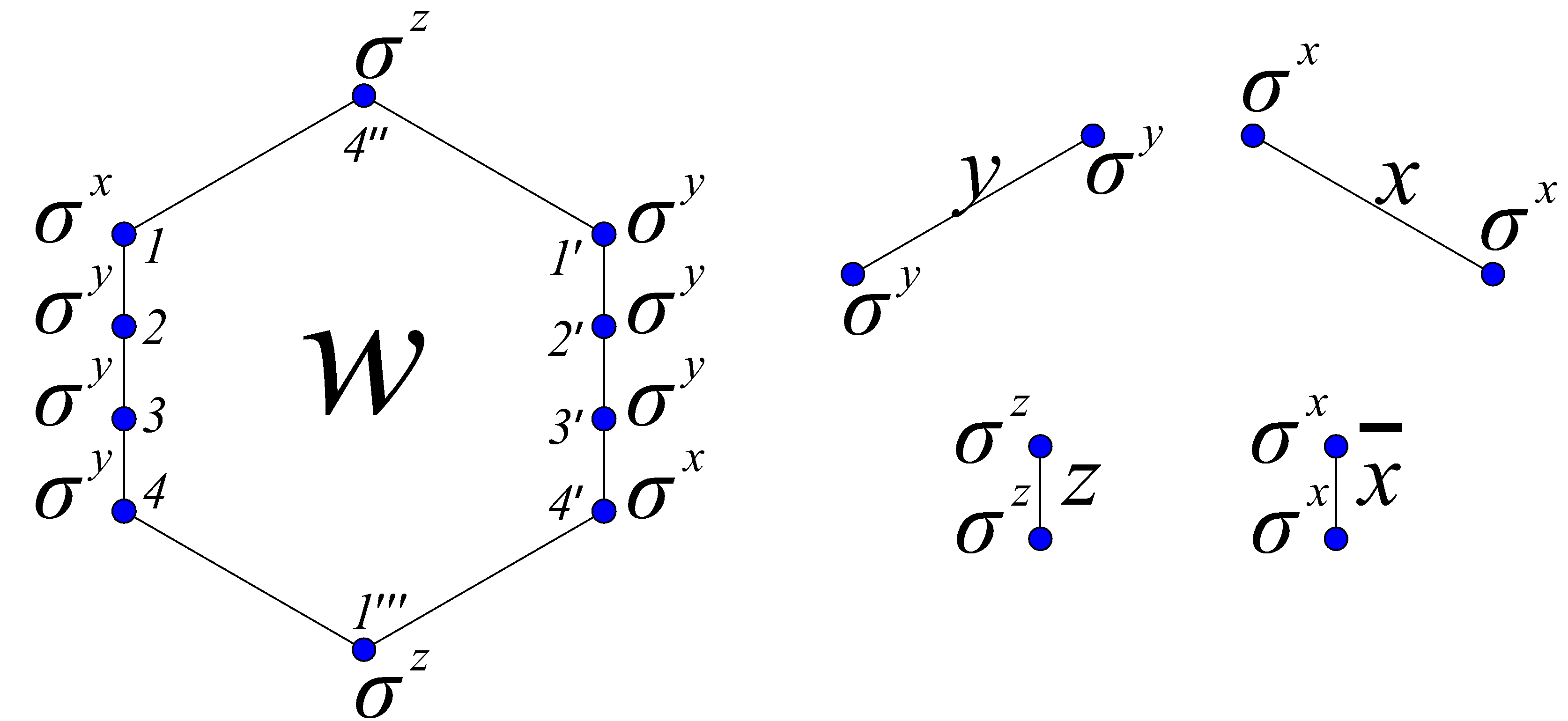}
\caption{(Color online)
  The spin sublattice defined on a honeycomb-like lattice with links of three directions ($x$, $y$ and~$z$) and $z$-link broken with $\bar{x}$-link;
  four types of links and the $w$-plaquette operator.
} \label{FigModelSpin}
\end{figure}

We focus on 2D model of spin subsystem interacting via the Kitaev interaction with a subsystem of spinless fermions.
The total Hamiltonian
\begin{equation}
{\cal H}={\cal H}_{s}+{\cal H}_{f}+{\cal H}_{int}
\label{eq:H}
\end{equation}
describes the spin and spinless fermion sublattices, and the interaction between them. The Hamiltonian of the spin subsystem ${\cal H}_{s}$ is defined on a honeycomb lattice with a \mbox{spin-$\frac{1}{2}$} on each site and an additional exchange interaction between spins within the $z$-links; it can be written in the framework of the Kitaev model~\cite{Kitaev} with an additional $\bar{x}$-links within the $z$-links
\begin{multline}
{\cal H}_{s} = \Delta_x \sum_{\nearsites{i,j}x} \sigma_i^x \sigma_j^x + \Delta_y \sum_{\nearsites{i,j}y} \sigma_i^y \sigma_j^y \\
{} + \Delta_z \sum_{\nearsites{i,j}z} \sigma_i^z \sigma_j^z + I \sum_{\nearsites{i,j}{\bar{x}}} \sigma_i^x \sigma_j^x,
\label{eq:Hs}
\end{multline}
where $\nearsites{i,j}$ is a pair of sites connected by $x$-, $y$-, $z$- and $\bar{x}$-links mentioned right after the pair notation,
$\sigma_{j}^\gamma$ are the three Pauli operators at a site $j$, and $\Delta_{\gamma}$ are the exchange integrals along all the links of corresponding direction \mbox{$\gamma=x,y,z$}. The interposed $\bar{x}$-link breaks $z$-link via the $I\sigma^x_i\sigma^x_j$ exchange interaction (the last term in (\ref{eq:Hs})) with the exchange integral $I$ between spins in two additional sites on the $z$-link (see Fig.~\ref{FigModelSpin}). For convenience, consider \mbox{$\Delta_\gamma, I > 0$}.

For the sublattice of the spinless fermions we will use the following tight-binding model that takes into account the interaction between nearest and next-nearest neighbors with equal hoppings and pairing amplitudes
\begin{multline}
{\cal H}_{f} = 
  - \ii t_1 \sum_{\nearsites{l,m}}\left( a^\dagger_{l} a^{\phantom{\dagger}}_{m\phantom{l}} + a_{l} a_{m} \right) \\ 
  - \ii t_2 \sum_{\nextnearsites{l,m}}\left( a^\dagger_{l} a^{\phantom{\dagger}}_{m\phantom{l}} + a_{l} a_{m} \right) + h.c.,
\label{eq:Hf}
\end{multline}
where $a_{l}^\dagger$ and $a^{\phantom{\dagger}}_{l}$ are the spinless creation and annihilation operators defined on the honeycomb fermion sublattice and satisfying the usual anticommutation relations, $t_1$ is an overlap integral of nearest-neighbor hopping between neighbor sites $\nearsites{l,m}$ from $l$ of the type 2 to $m$ of the type 3, $t_2$ is a clockwise (relatively a honeycomb) next-nearest-hopping between second-neighbors $\nextnearsites{l,m}$, $l$ and $m$ range all sites of both type $2$ and $3$.

\begin{figure}[tbp]
\centering{\leavevmode}
\includegraphics[width=7cm]{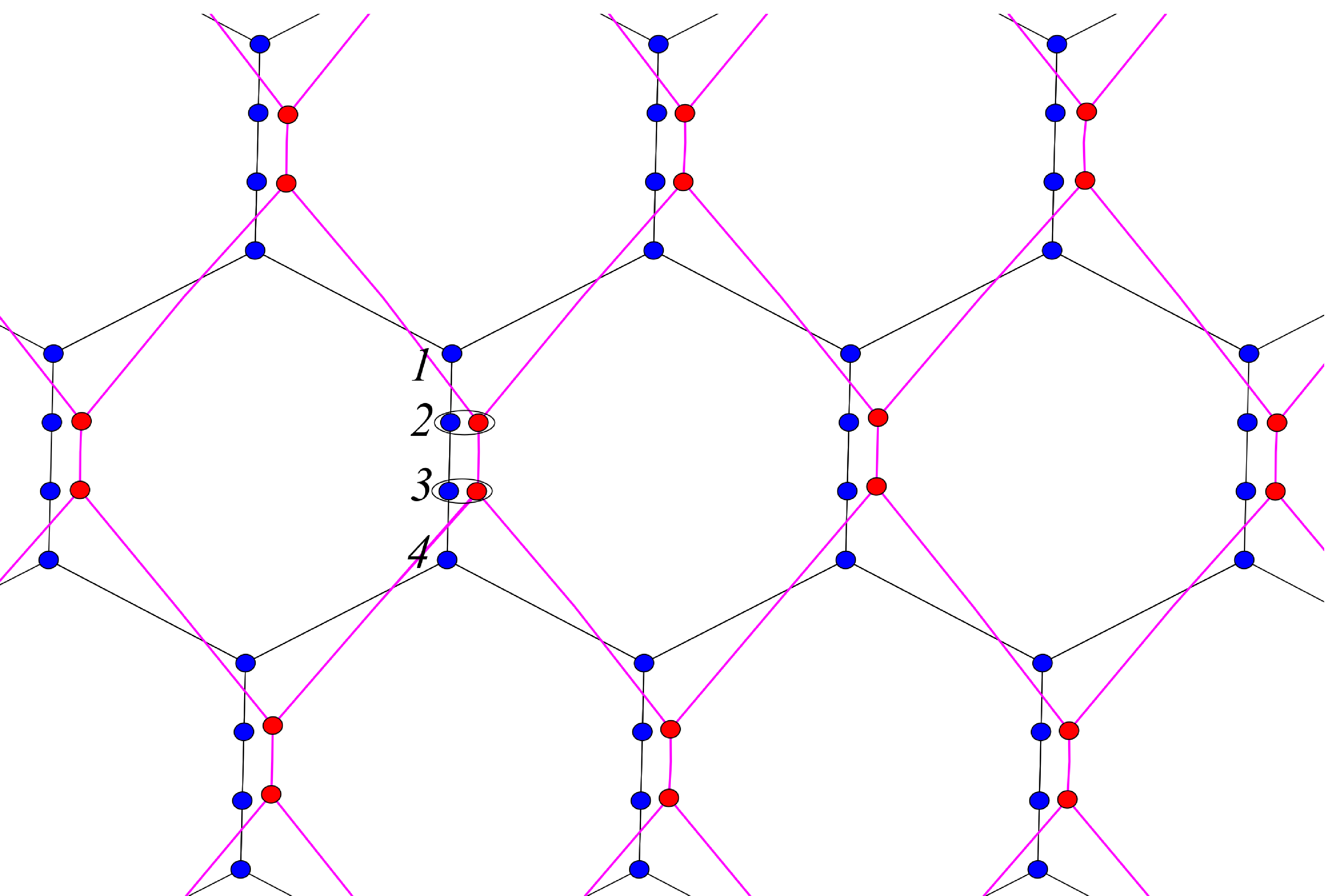}
\caption{(Color online)
  The topological Kondo lattice consisting of the spin (blue circles) and fermion (red circles) sublattices.
  Red circles form a honeycomb lattice up to a continuous transformation.
  The interaction between sublattices is defined on connected red-blue pairs.
} \label{FigModel}
\end{figure}

The unit cell of the lattice consists of four sites of spin subsystem (namely \mbox{$1-4$} in Fig.~\ref{FigModel}) and two sites of fermion subsystem (marked by $2$ and $3$). The term ${\cal H}_{int}$ is governed by a contact interaction between itinerant states of spinless fermions and spin operators at the sites \mbox{$2-3$} that are interposed into the $z$-links
\begin{equation}
{\cal H}_{int} = \lambda \sum_{l} (2 n_{l}-1)\sigma^y_{l},
\label{eq:Hint}
\end{equation}
where $\lambda$ is the coupling parameter, $n_l=a^\dagger_{l}a^{\phantom{\dagger}}_{l}$, $l$ ranges all pairs of spin and fermion sites, namely $2$ and $3$ of each cell (see Fig.~\ref{FigModel}).

The Hamiltonian~(\ref{eq:H}) defines an exactly solvable model of topological Kondo insulator on a honeycomb topological Kondo lattice.
The Hamiltonian of the spin sublattice ${\cal H}_s$ can be exactly diagonalized using the representation of the Pauli operators in terms of a related set of Majorana fermions $b_j^\gamma$ and $c_j$ with commutation rules
$$\{c_i,c_j\}=2\delta_{ij}, \quad \{b^\gamma_i,b^{\gamma'}_j\}=2\delta_{\gamma\gamma'}\delta_{ij}, \quad \{c_i,b^\gamma_j\}=0,$$
with the substitution $\sigma^\gamma_j = \ii b^\gamma_j c_j$~\cite{Kitaev} ($\gamma=x,y,z$).
We introduce two types of Majorana fermions on each site for the sublattice of spinless fermions
$$d_l = a^{\phantom{\dagger}}_{l} + a^\dagger_{l} \quad\textrm{ and }\quad g_l = \frac{a^{\phantom{\dagger}}_{l}-a^\dagger_{l}}{\ii},$$
the Hamiltonian $\cal H$ can thus be rewritten in the form
\begin{multline}
{\cal H} = -\ii \sum_{\beta=x,y,z,\bar{x}}\sum_{\nearsites{i,j}\beta} A_{ij}^\beta c_i c_j- \ii t_1 \sum_{\nearsites{l,m}} d_l d_m
  \\-\ii t_2 \sum_{\nextnearsites{l,m}} \nu_{lm} d_l d_m + \lambda\sum_{l} g_l d_l c_l b_l^y,
\label{eq:HMaj}
\end{multline}
where the matrix $A$ consists of $A_{ij}^\gamma = \Delta_\gamma u_{ij}^\gamma$ for the directed links $\gamma=x,y,z$ and $A_{ij}^{\bar x}=Iu_{ij}^x$ for the intercalated $\bar{x}$-link, $u_{ij}^\gamma = -u_{ji}^\gamma = \ii b_{i}^\gamma b_{j}^\gamma$ and $\nu_{ij}=\pm1$ stands for clockwise (anticlockwise) next-near-neighbour hopping inside corresponding plaquette.

The physical subspace is defined by the constrain $D_j\vert\psi\rangle_{\textrm{phys}}=\vert\psi\rangle_{\textrm{phys}}$ with $D_j=b^x_jb^y_jb^z_jc_j$. The operator $D_j$ acts as the identity operator on the physical subspace~\cite{Kitaev}, it commutes with the Hamiltonian (\ref{eq:HMaj}).

The plaquette operator~\cite{Kitaev}
\mbox{$w_s=\sigma^x_{s1}\sigma^y_{s2}\sigma^y_{3}\sigma^y_{s4}\sigma^z_{s1'''} \sigma^x_{s4'}\sigma^y_{s3'}\sigma^y_{s2'}\sigma^y_{s1'}\sigma^z_{s4''}$}
is defined by a product of the $ u_{ij}^{\gamma}$ operators around a plaquette $s$, see Fig.~\ref{FigModelSpin}.
Operators $u_{ij}^{\beta}$ are constants of motion with eigenvalues $\pm 1$. Each plaquette operator $w_s$ has thus two eigenvalues $\pm 1$ and it is interpreted as a magnetic flux through the plaquette $s$. The operators $u_l=\ii g_{l}^{\phantom{y}}b_{l}^y$ are the constants of motion with eigenvalues $\pm 1$. Thus the Hilbert space of states might be split into eigenspaces, where all operators $u_{ij}^{\gamma}$ and $u_l$ might be replaced with their eigenvalues. The variables $u_{ij}^{\gamma}$ and $u_l$ are identified with a static ${\mathbb{Z}}_2$ gauge fields on the bonds. The Hamiltonian~(\ref{eq:HMaj}) is thus reduced to a quadratic form. The itinerant fermions and localized moments reconcile with each other via the hybridization  with pairing states of spin-fermion Majorana fermions on the lattice sites.

To solve the model exactly, one converts each vortex sector to free spinless fermions~\cite{Kitaev}.
Numerically, we have computed the ground-state energy of the model for a set of
finite size systems and for various set of the exchange integrals. In all cases
we found that the ground-state energy is minimized by the same
uniform flux pattern. The vortex uniform sector with $w_s=1$ for all plaquette 
operators $w_s$ is the ground state of the model~\cite{Kitaev, Lieb}. The model is solved
analytically for the uniform configurations, due to the translational
invariance of the lattice. We now focus on the vortex-free configuration ($w_s = 1$) for the
Hamiltonian ${\cal H}$ of the entire system~(\ref{eq:H}).
A contact interaction~(\ref{eq:Hint}) breaks both particle-hole symmetry and time reversal (TR) symmetry of the model Hamiltonian and gives a nontrivial ground-state phase diagram of the system.

\section{Phase analysis}

\subsection{Noninteracting subsystems}

\begin{figure}[t!]
\centering{\leavevmode}
\includegraphics[width=5cm]{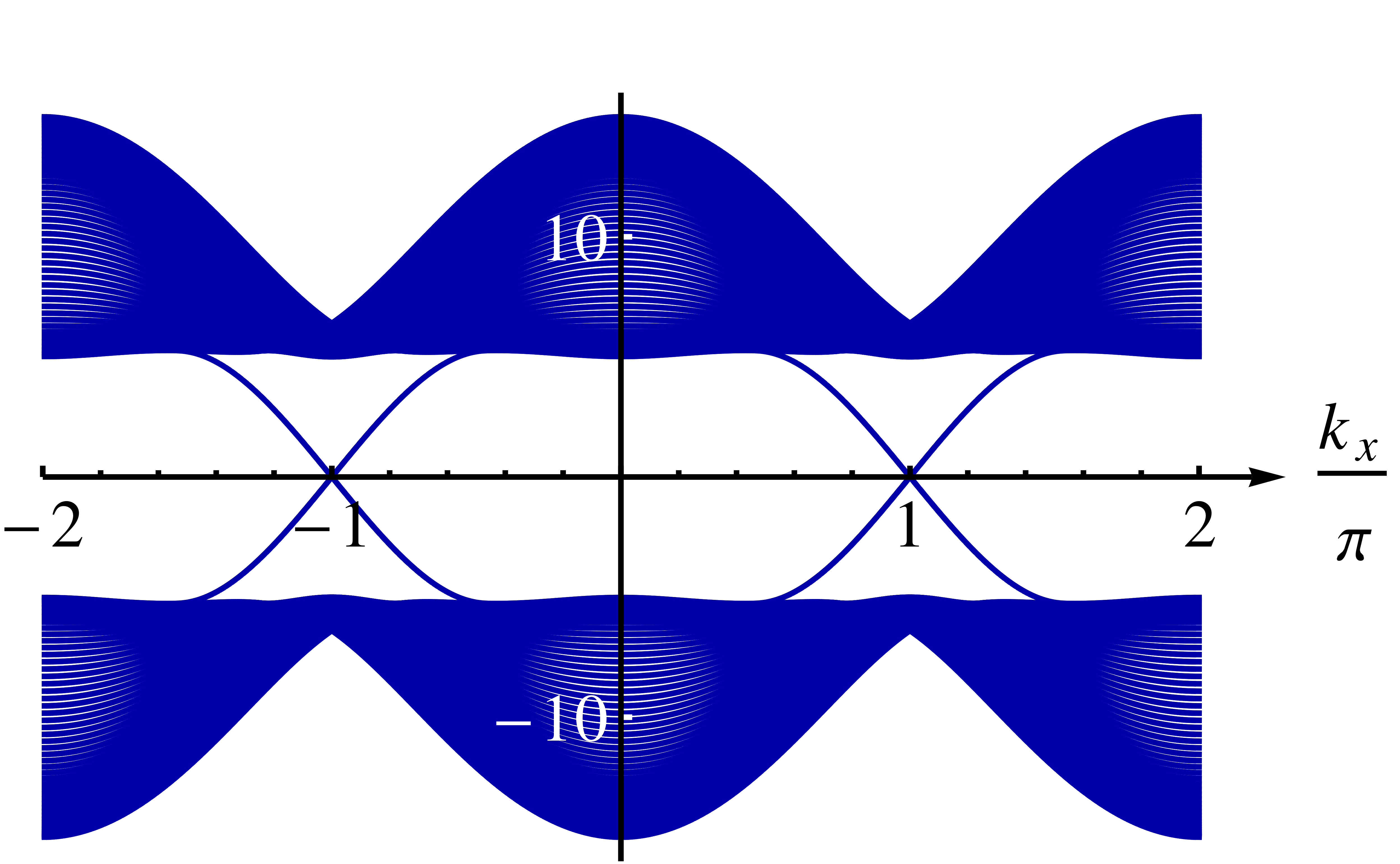}
\caption{(Color online)
    The energy levels of the noninteracting ($\lambda=0$) fermion subsystem calculated on a cylinder with open zig-zag boundary conditions, $t_1=5$, $t_2=1$.
} \label{NoninteractingFermionStates}
\end{figure}

In the absence of the interaction~(\ref{eq:Hint}) the system is a sum of two decoupled spin and fermion subsystems.
Spinless fermion subsystem is always in topologically nontrivial phase if $t_1\neq 0$ and $t_2\neq 0$~\cite{Hal,k1},
its spectrum is gapped, one chiral edge mode is presented, see Fig.~\ref{NoninteractingFermionStates}.
The phase is characterized by Chern number $C_{ch}$ of the charge sector, $C_{ch}=\pm 1$ depending on the sign of $t_1/t_2$.

\begin{figure}[t!]
\centering{\leavevmode}
\includegraphics[width=6cm]{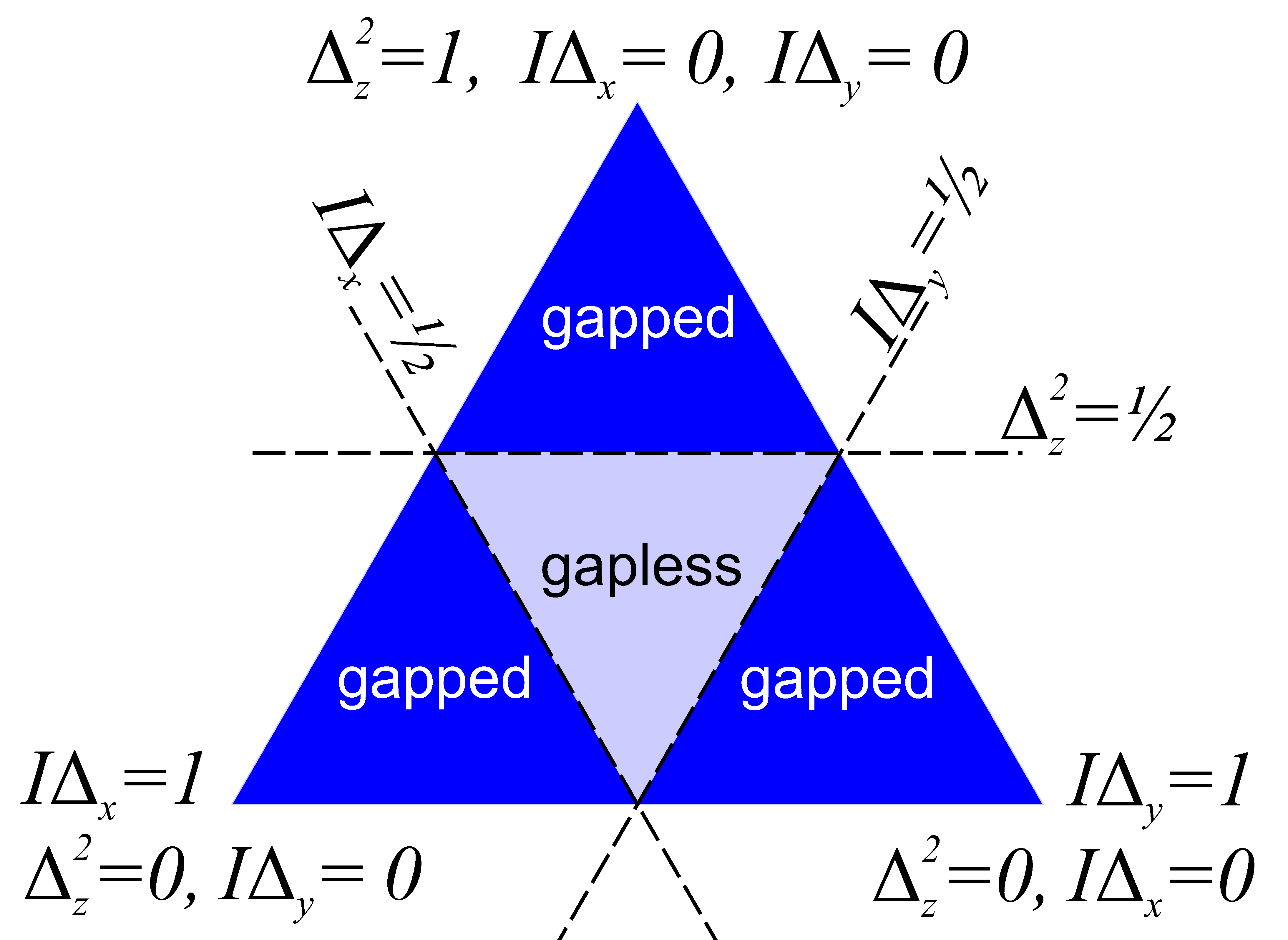}
\caption{(Color online)
Phase diagram of Kitaev's type for positive exchange integrals that satisfy $I\Delta_x+ I\Delta_y+\Delta_z^2=1$.} \label{Diag}
\end{figure}

Topological phases of noninteracting spin subsystem are also associated with Chern number $C_s$.
It is convenient to illustrate the ground-state phase diagram of the spin subsystem using Kitaev's diagram~\cite{Kitaev} Fig.~\ref{Diag} --- a section by the plane \mbox{$\Delta_z^2 +I\Delta_x +I\Delta_y = \operatorname{const}$} in coordinates $I\Delta_x$, $I\Delta_y$ and $\Delta_z^2$. There are two topological distinct phases: topological trivial phase (with gapped states) indexed by $C_s=0$ and topological nontrivial phase (with gapless states) with $C_s=1$ are separated by lines of quantum phase transitions. In the case of noninteracting subsystems, an external magnetic field breaks TR symmetry of spin subsystem and opens a gap in the spectrum of the Majorana fermions~\cite{Kitaev} in the gapless region. The structure of energy levels with edge states in the spin sector is shown on Figs~\ref{NoninteractingStates} for both phases.

\revision{
\begin{figure}[t!]
\centering{\leavevmode}
\includegraphics[width=4cm]{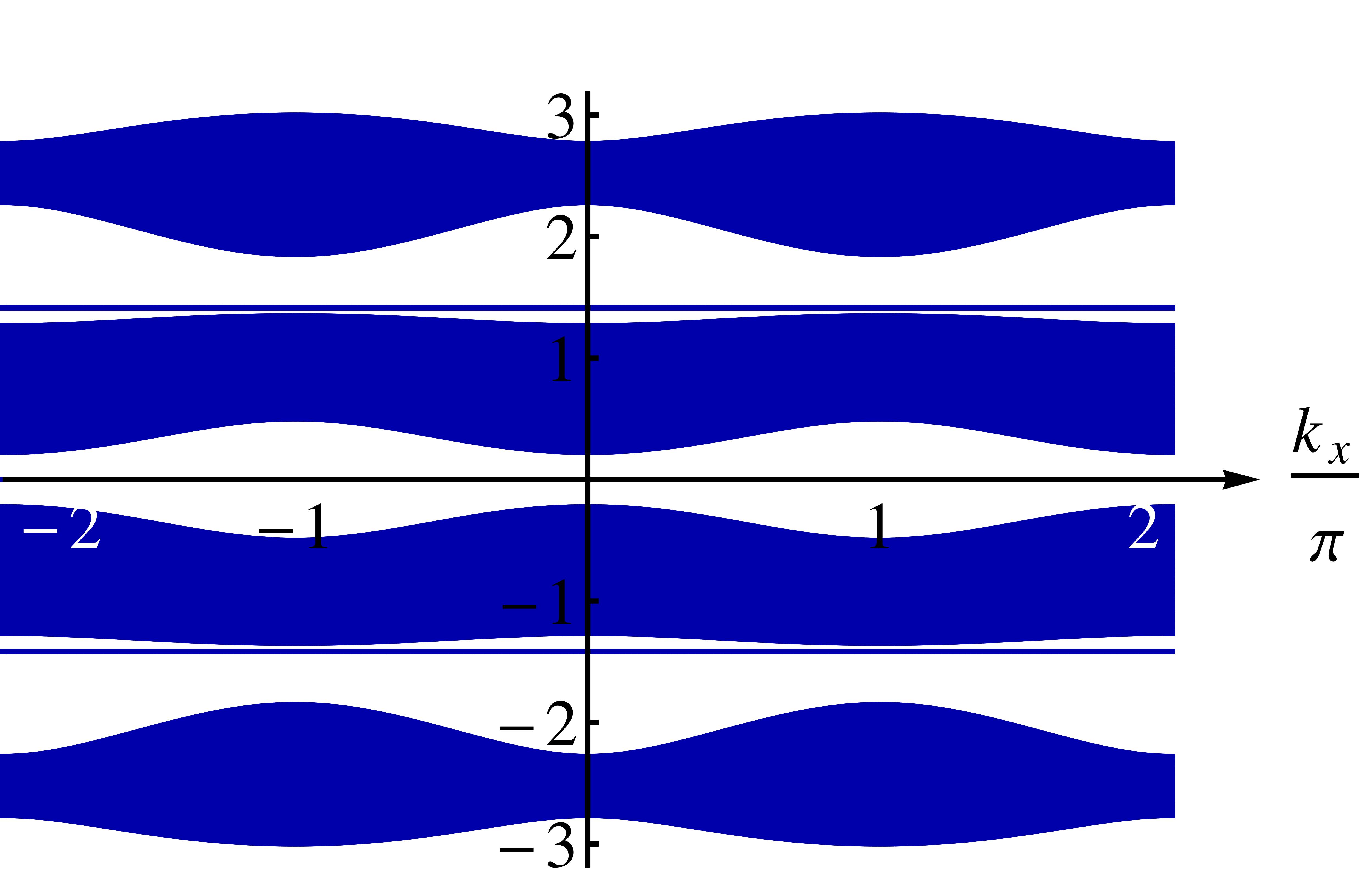}
\includegraphics[width=4cm]{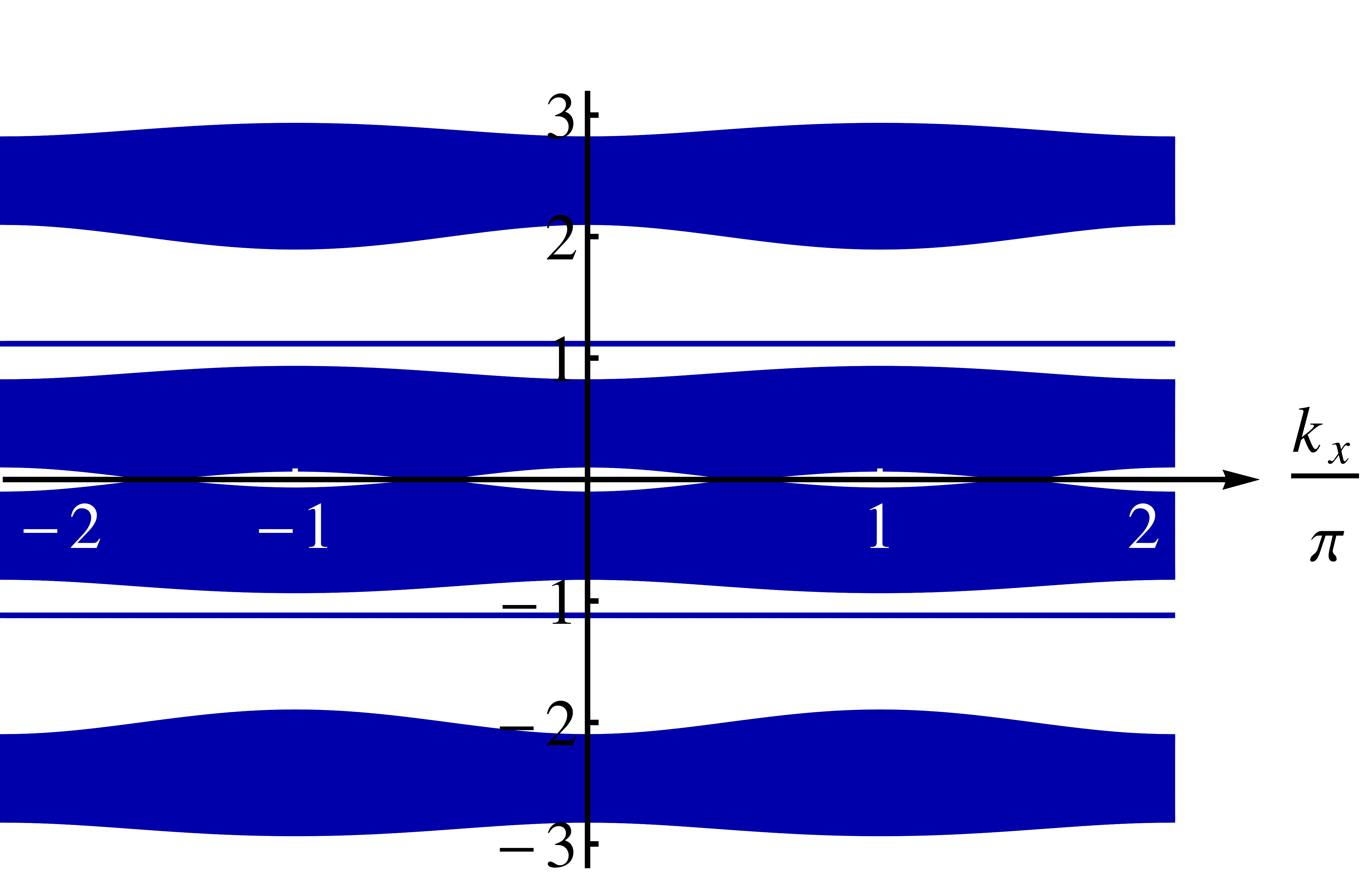}
\caption{(Color online)
    The energy levels of the noninteracting ($\lambda=0$) spin subsystem calculated on a cylinder with open boundary conditions for zig-zag boundary along $x$-direction, gapped $I=1$ (left) and gapless $I=\frac12$ (right) case at $\Delta_z=1$, $\Delta_x=\frac{1}{2}$, $\Delta_y=2$.
} \label{NoninteractingStates}
\end{figure}
}

\subsection{Phase diagram of interacting subsystems}

\begin{figure}[t]
\centering{\leavevmode}
\includegraphics[width=6cm]{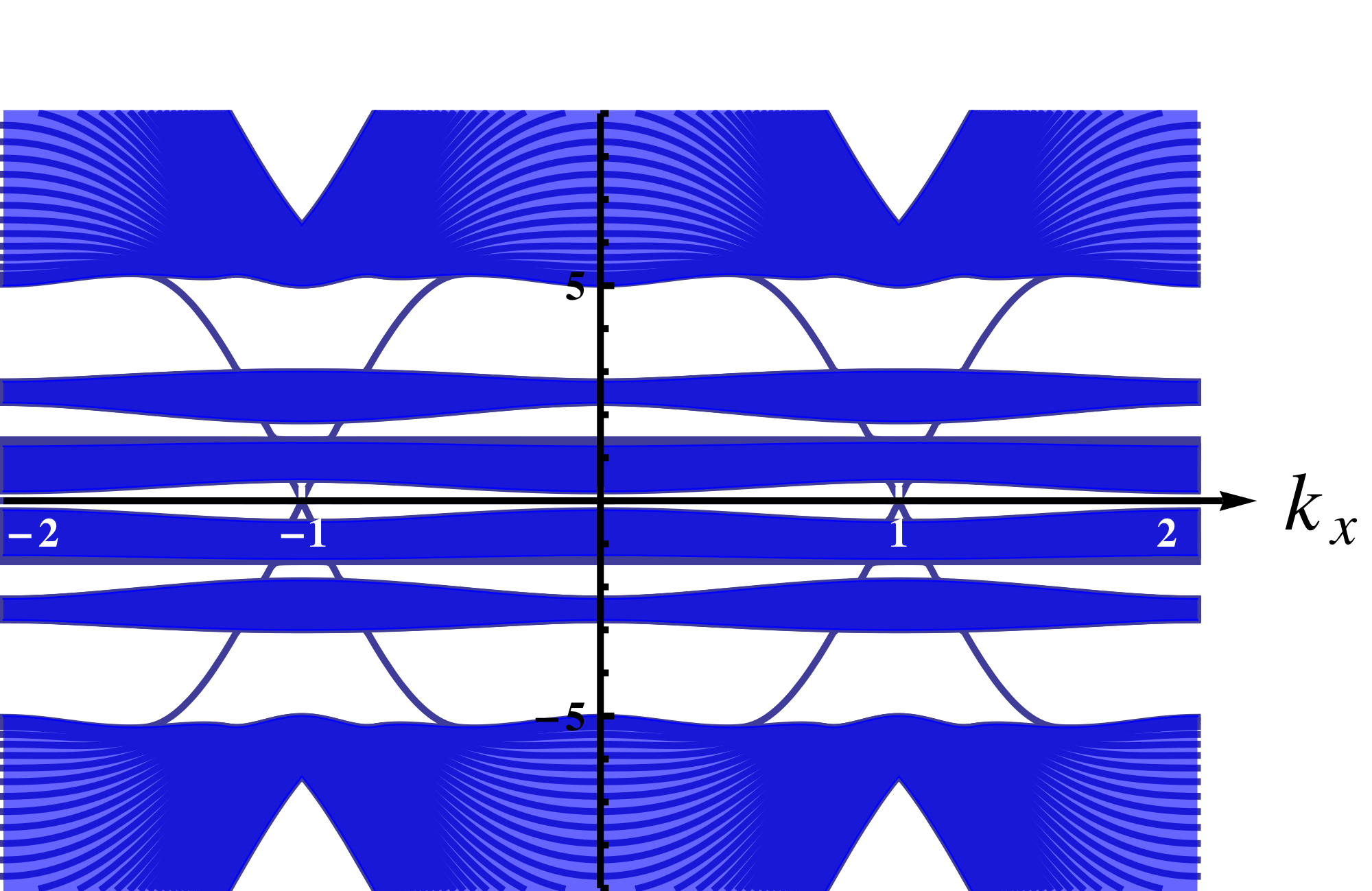}
\caption{(Color online)
    The energy levels of the system calculated on a cylinder with open boundary conditions for zig-zag boundary along $x$-direction 
    with $t_1=5$, $t_2=1$, $\Delta_z=1$, $\Delta_x=\frac{1}{2}$, $\Delta_y=2$, $I=1$ and $\lambda=\frac{1}{10}$.
} \label{Spectrum}
\end{figure}

\begin{figure}[t!]
\centering{\leavevmode}
\includegraphics[width=6cm]{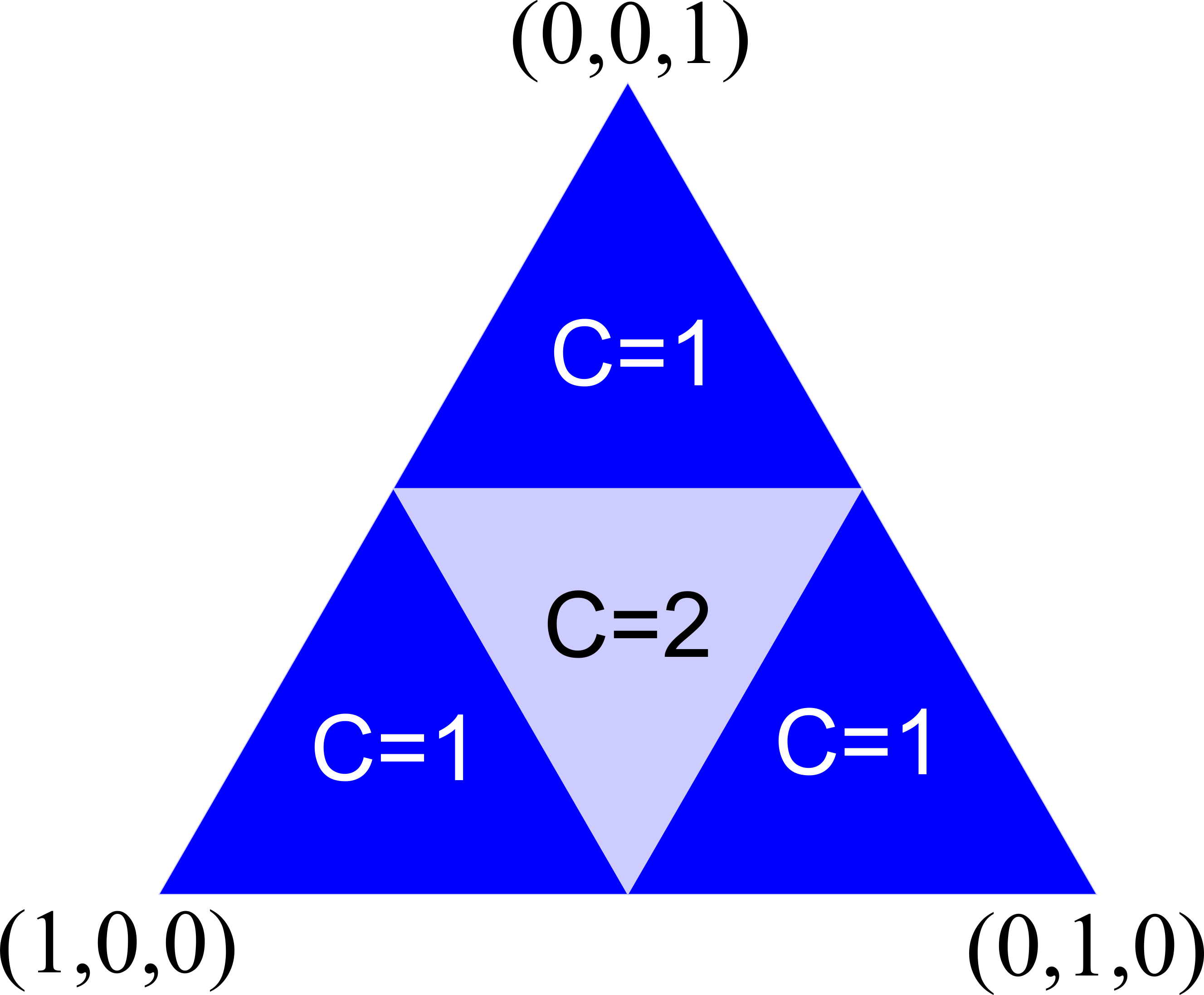}
\caption{(Color online)
  Phase diagram in coordinates $(I\Delta_x, I\Delta_y, \Delta_z^2)$ with a exchange intergal satisfying a normalization condition $I\Delta_x+I\Delta_y+\Delta_z^2=1$ for arbitrary $t_1$, $t_2$ and $\lambda$.} \label{Diag2}
\end{figure}

Let us consider the adiabatic connection of the subsystems via the interaction~(\ref{eq:Hint}) and evolution of the ground state of the system along $\lambda$ and $I$ directions. We use the fixed parameters of the Hamiltonian~(\ref{eq:H}) $t_1=5$ and $t_2 = 1$ in the case of an anisotropic spin-exchange interaction $\Delta_x=\frac{1}{2}$, $\Delta_y=2$ and $\Delta_z=1$ to demonstrate a band structure and edge modes observed in common case.
The spectrum of a noninteracting fermion subsystem is gapped,
a noninteracting spin subsystem has gapless spectrum of the Majorana fermions when $\frac{2}{5}\leq|I|\leq\frac{2}{3}$, otherwise it is gapped. There are six bands, corresponding to six sites per unit cell: two high energy fermion bands and four low energy spin bands (see Fig.~\ref{Spectrum}).
The structure of the edge states depends on a direction of a boundary. We consider the zig-zag boundaries along $x$- and $y$- exchange bonds for illustration.
The direct interaction (\ref{eq:Hint}) between subsystems breaks TR symmetry of (\ref{eq:H}) and opens the hybridization gap in the spectrum of spin-fermion excitations at the half-filling.
The interaction term (\ref{eq:Hint}) breaks TR symmetry of the complex system, the coupling constant $\lambda$ `works' as an external magnetic field in the Kitaev model, it opens the gap in the gapless topological spin phase. The structure of the spectrum and the topological number of the entire system do not vary in the process of the interaction's switch and the evolution at the adiabatic connection when the value of $\lambda$ increases. The Chern number of the system (\ref{eq:H}) for arbitrary $\lambda$ depends on the Chern numbers of the spin and charge sectors for $\lambda=0$. There are two possible scenarios of the formation of the Kondo insulator phase state along the $\lambda$ direction depending of the entry conditions at $\lambda=0$:
\begin{itemize}
\item the spin subsystem is in topological trivial phase (gapped states), the fermion subsystem is in the state of the topological insulator;
\item both the spin and fermion subsystems are in topological nontrivial phases.
\end{itemize}
Thus the phase diagram of the model system in the coordinates $I\Delta_x$, $I\Delta_y$, $I\Delta_z$ and $\lambda$ is similar to the one in Fig.~\ref{Diag}, where the phase diagram of spin subsystem lies in the base of a prism and the $\lambda$ axis is normal to the base (see Fig.~\ref{Diag2}). The planes of the topological phase transitions, in which the bulk gap vanishes in the spin sector at $\lambda=0$, separate the topological phases with the Chern numbers 1 and 2, as it will be shown below.
We consider the cases $I=1$ and $I=\frac{1}{2}$ to illustrate these two scenarios of behavior and investigate peculiarities of the topological states.

\begin{figure}[t]
\centering{\leavevmode}
\includegraphics[height=3cm]{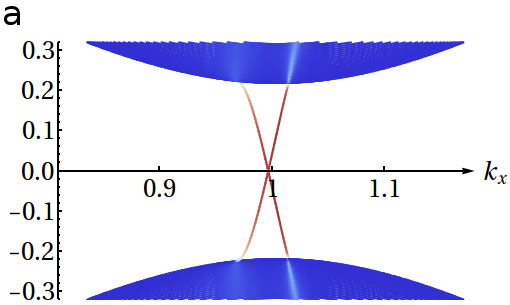} \includegraphics[height=2.5cm]{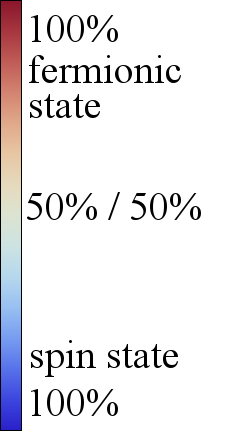}
\includegraphics[width=4cm]{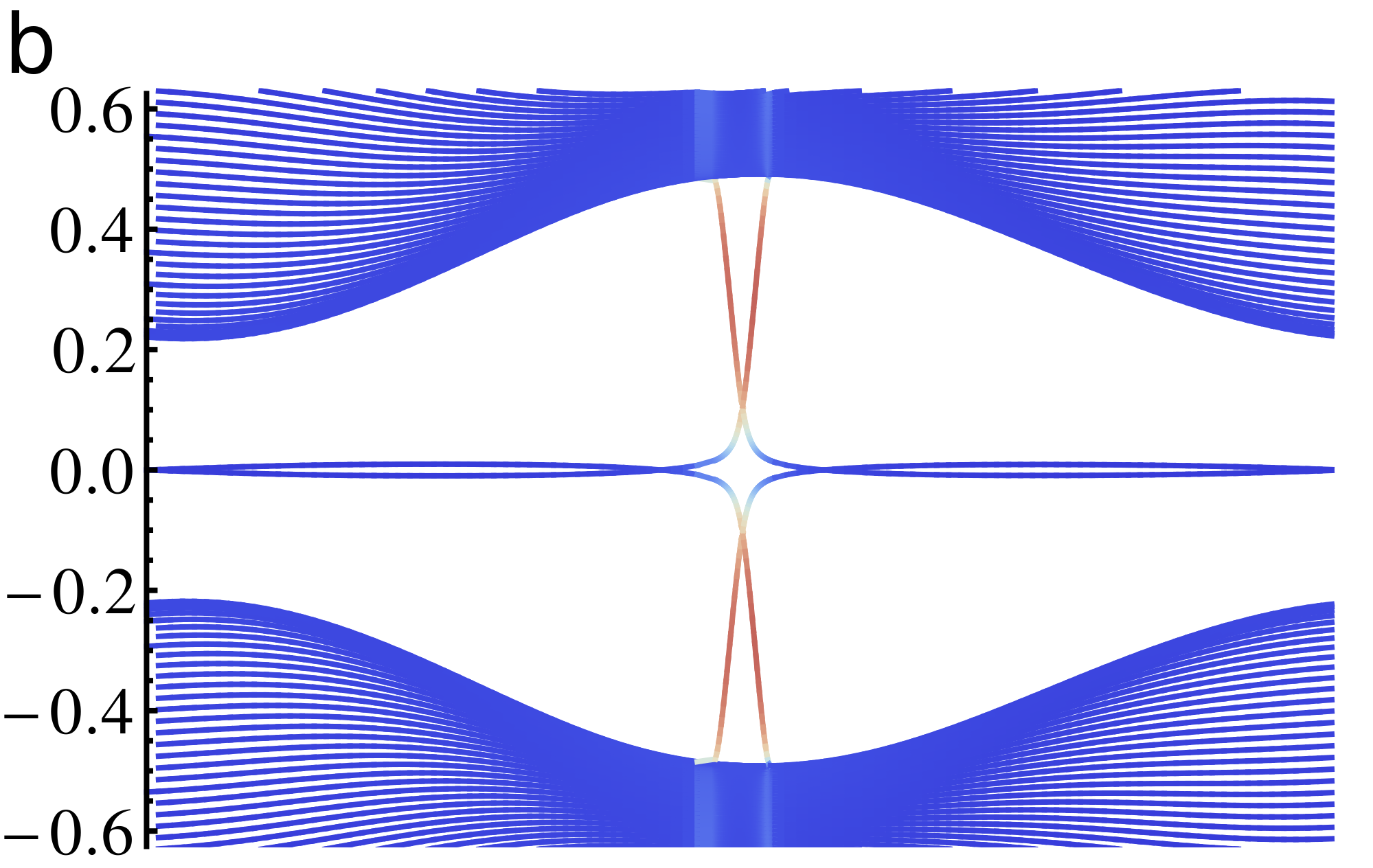} \includegraphics[width=4cm]{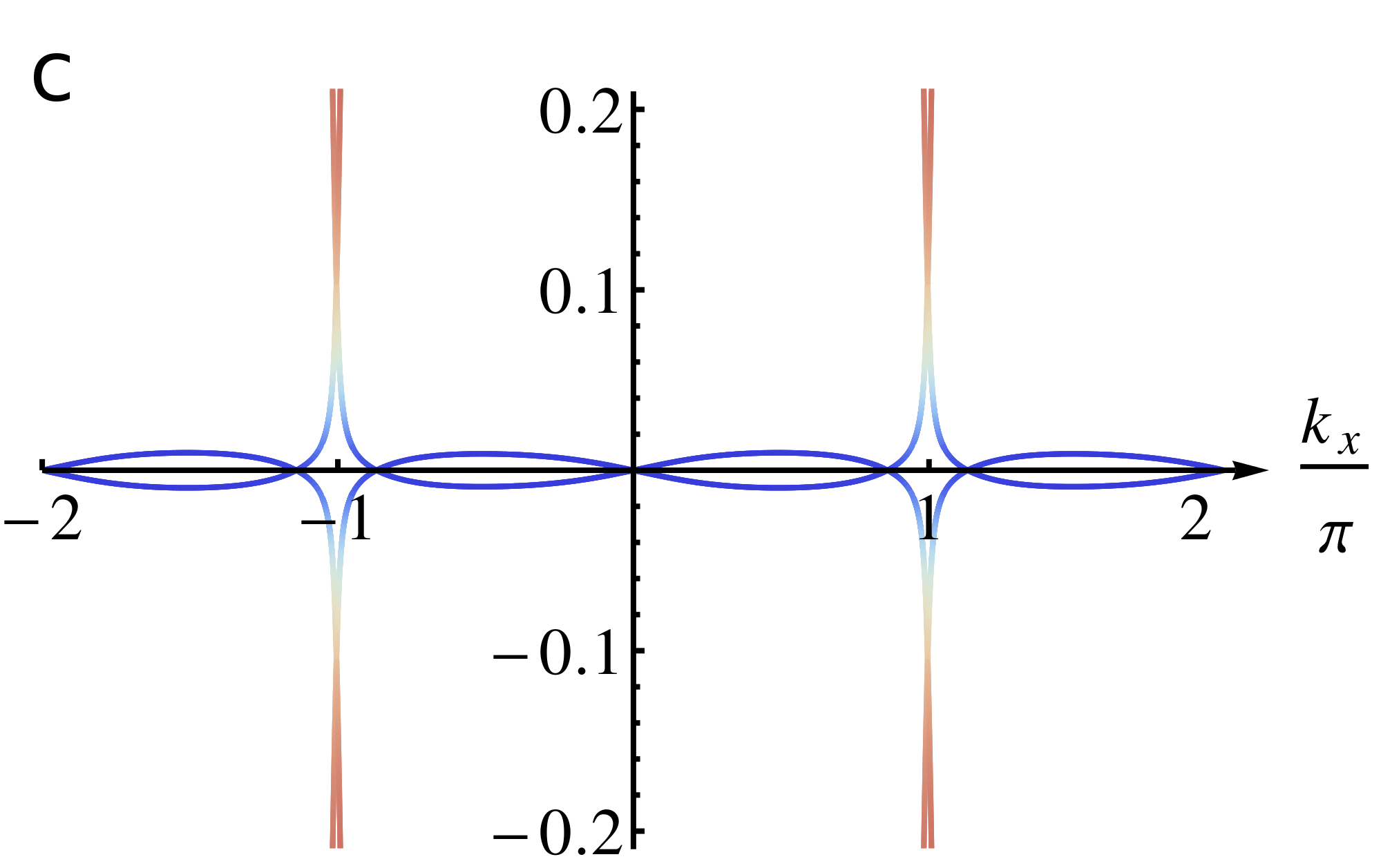}
\caption{(Color online)
    A low energy brushed spectrum (as in Fig.\ref{Spectrum})
    at $t_1=5$, $t_2=1$, $I=1$, $\Delta_x=\frac{1}{2}$, $\Delta_y = 2$, $\Delta_z=1$ with boundary along $y$-direction and a hybridization $\lambda=\frac{1}{5}$ a) and along $x$-direction and a hybridization $\lambda=\frac{1}{2}$ b-c). Each point has its color depending on fermionic rate of its state which is calculated as a projection to the neat fermion state subspace $\langle \psi | \operatorname{pr}_{\mathrm{ferm}} | \psi \rangle$. A color scale is given.
} \label{Edge1}
\end{figure}

\subsection{Scenario 1}

At $I=1$ and $\lambda \neq 0$ the phase state of the system is characterized by two (spin  and fermion) gaps in the excitation spectrum. Calculations of the Chern number and edge states show that  $C_{ch}=1$ and $C_{s}=0$ at $\lambda=0$, a total Chern number is equal to 1 for arbitrary values of $\lambda$.
The structure of edge states depends on a direction of folding: $y$-direction boundary (Fig.~\ref{Edge1}a) carries chiral gapless edge modes specified by fermion sublattice and thus all low-level excitations are fermionic, while $x$-direction boundary (Fig.~\ref{Edge1}b) performs chiral edge modes crossing the Fermi level three times with excitations of spin type (Fig.~\ref{Edge1}c). These edge modes are associated with $C=1$, but topological insulator's behaviors are different: it acts as fermion topological insulator in $y$-direction and as spin topological insulator in $x$-direction.

\begin{figure}[t]
\centering{\leavevmode}
\includegraphics[height=3cm]{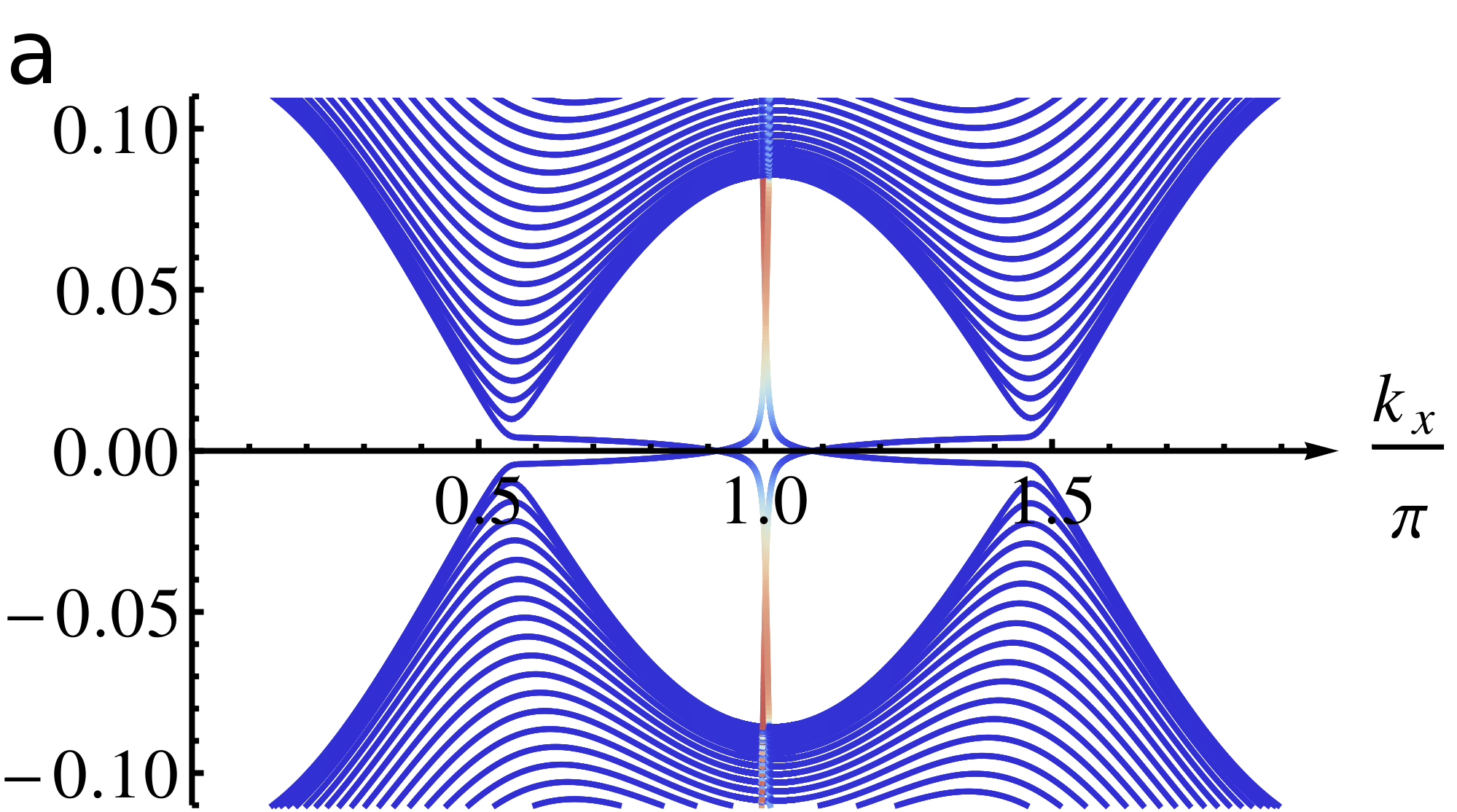} \includegraphics[height=2.5cm]{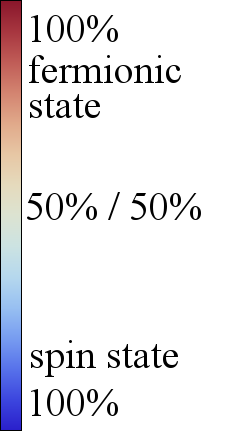}
\includegraphics[width=4cm]{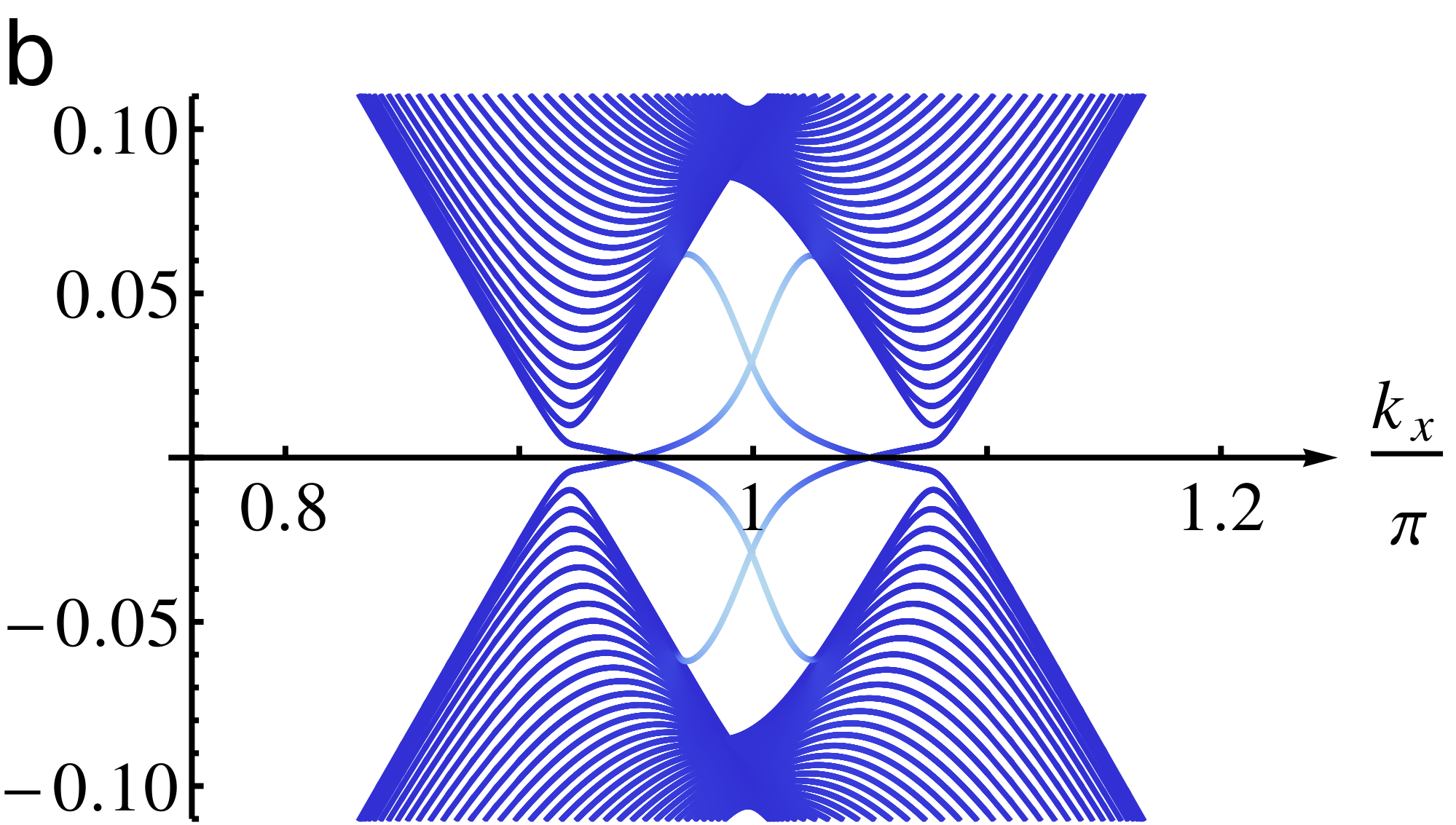} \includegraphics[width=4cm]{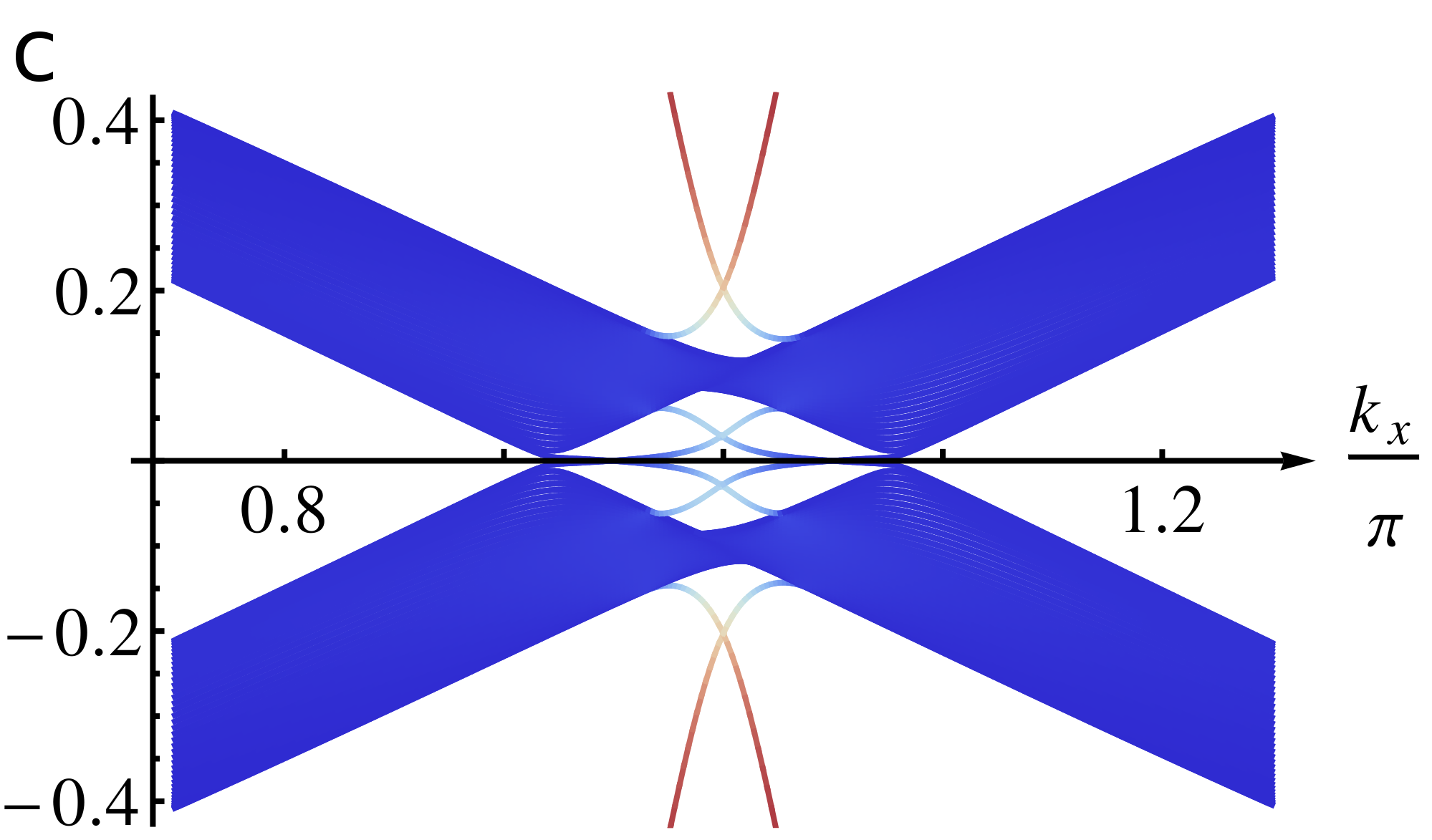}
\caption{(Color online)
  A low energy brushed spectrum (as in Fig.\ref{Edge1}) at $t_1=5$, $t_2=1$, $I=\frac{1}{2}$, $\lambda=\frac{1}{5}$; $\Delta_x=\frac{1}{2}$, $\Delta_y = 2$, $\Delta_z=1$ with boundary along $x$-direction a) and $y$-direction b-c). A color scale is given.
} \label{Edge2}
\end{figure}

The interaction term (\ref{eq:Hint}) also directly hybridizes the edge states, therefore the edge states are a result of total spectrum of the system on the one hand, and they are a result of direct hybridized interaction on the boundary on the other hand.

\subsection{Scenario 2}

Now consider the second scenario of behavior of the Kondo lattice when the closed at $\lambda=0$ gap in the spectrum of spin excitations opens for $\lambda \neq 0$, as it is illustrated at $I=\frac{1}{2}$ in Fig.~\ref{Edge2}.
At $\lambda=0$ a spin gap closing makes the topological index $C_{s}$ ill-defined. Taking into account a magnetic field as a week disturbance that breaks TR symmetry of spin subsystem and opens the gap in the spin sector of the spectrum of Majorana fermions~\cite{Kitaev}, we calculate $C_s$ and edge states of the spin subsystem. According to obtained calculations, $C_s=1$ (the same value as $C_{ch}$).
For $\lambda \neq 0$, the gap opens due to a strong hybridization in a low energy region of the spectrum even in the absence of an external magnetic field. The total Chern number of the system is equal to two, there are two gapless chiral edge modes (hybridized quasi spin and fermion ones), see Fig~\ref{Edge2}. A low energy behavior of the edge modes has a spin nature, as we see in Figs~\ref{Edge2}, and as result, a topological spin insulator state with two quasi spin edge modes is realized in the Kondo lattice in both directions of bounds.

\subsection{Overall notes}

The model of the proposed 2D topological Kondo lattice takes into account an interaction between different topological subsystems, namely, spin and fermion lattices. The spin-fermion interaction leads to the hybridization of the Majorana fermions of the subsystems. The considered interaction is reduced to the hybridization between excitations in the spin and charge sectors, as result  the Chern number of the system is characterized by one (total) Chern number, that depends on spin and charge Chern numbers of the corresponding noninteracting subsystems. 

In spite of the Hamiltonian (\ref{eq:HMaj}) does not conserve the total number of particles, the Fermi surface is sharply defined. The hybridization of itinerant spinless fermions and local moments form the volume of the Fermi sea
$\int_{BZ} d\mathbf{k} = n + 1,$ where $n$ is a density of fermion and integrating is over the Brillouin zone of the system.
A stability of the phase of the Kondo insulator does not depend on the eigenvalues of the local operators $u_l = 2\mathfrak{n}_l - 1$ because the ground state energy is realized both $\mathfrak{n}_l=1$ and  $\mathfrak{n}_l=0$, where $\mathfrak{n}_l$ is a number of fermions on the site $l$ of the fermion sublattice. Therefore there are free bulk states of Majorana fermions (with zero energy) located at the sites of the fermion sublattice.
The total ground-state energy of the system reaches minimum at the half-filling of fermion and spin subsystems $n=1$, this phase state is also realized under an arbitrary doping $p=n-1$ of particles, because the rest of $p$ particles are located at the fermion sublattice (on the sites with $\mathfrak{n}_l =1$) and occupy the Fermi surface $\varepsilon_F=0$ in the insulator phase. The total energy of the system remains the same, it does not depend on the doping of the system, a static $\mathbb{Z}_2$ gauge field `works' as a reservoir for doped particles. In contrast to traditional insulator phase, the Fermi surface is occupied and an insulator gap is effectively two times less.

\section{Conclusions}

In summary, we have considered the implication of the adiabatic connection between spin and fermion subsystem defined on a honeycomb Kondo lattice.
There exist two possibilities of realization of the Kondo insulator state, both cases have been considered. We have calculated a hybridization gap in the framework of the  model proposed (up to now the mechanism of forming of the hybridization gap in the state of the Kondo insulator has been unknown). The spinless fermions are localized at the lattice sites due to the contact interaction with local moments, this phenomenon is analogous to the Kondo screening in real space (on lattice sites). The itinerant states of spinless fermions hybridize with spin excitations, as a result, the gap opens at the half-filling. The gapless edge states form a surface subband of chiral Majorana fermions. It is shown that hybridization within topological Kondo insulator can lead to changing fermionic topological insulator into spin one in sense of low energy edge excitations.

In the spin-wave approach, the spin excitations are bosons, they are not hybridized with fermions. The ground state of the Kondo lattice at half-filling can be explained by a hybridization gap that arises from a hybridization of itinerant electrons and local moments. In this letter we have firstly proposed the mechanism of hybridization between local spins and  itinerant spinless fermions in the framework of the proposed exactly solvable model.

\end{document}